\documentclass[12pt]{iopart}

\usepackage{graphicx}
\usepackage{iopams}
\usepackage{bm}
\usepackage{subfigure}

\begin{document}


\title{Casimir Effect in $E^3$ closed spaces}

\author{Mariana P. Lima}
\address{Instituto de F\'\i sica, UnB \\ 
Campus Universit\'ario Darcy Ribeiro\\ 
Cxp 04455, 70919-970, Bras\'\i lia DF \\
Brasil 
}%
\ead{mariana@unb.br}
\author{Daniel M\"uller}
\address{Instituto de F\'\i sica, UnB \\ 
Campus Universit\'ario Darcy Ribeiro\\ 
Cxp 04455, 70919-970, Bras\'\i lia DF \\
Brasil 
}%
\ead{muller@fis.unb.br}
\date{\today}

\begin{abstract}
As it is well known the topology of space is not totally determined by 
Einstein's equations. It is considered a massless
scalar quantum field in a static Euclidean space of dimension $3$. The 
expectation value for the energy density in all compact orientable 
Euclidean $3$-spaces are obtained in this work as a finite summation
of Epstein type zeta functions. The Casimir energy density for these 
particular manifolds is independent of the type of coupling with curvature. 
A numerical plot of the result inside each Dirichlet region is obtained.

\end{abstract}
\pacs{04.62.+v; 03.70.+k}
                            
\maketitle

\section{Introduction} 

The Universe today is very close to homogeneous and isotropic. 
Then the most important 3-Manifolds
in connection with cosmology are $S^{3},\: H^{3}$ and $E^{3}.$ 
It is well known that $R\times S^{3},\: R\times H^{3}$ and $R\times E^{3}$
are solutions of Einstein's equations (EQ). It is irrelevant for the EQ, if the
spatial sections have a non trivial topology modelled on them, or not. 

There are many effects that arise if space is compact. The first astrophysical
limits on the topology of the universe were obtained for a 3-torus
$T^{3}$. Accordance with the homogeneity of the cosmic microwave background
radiation CMBR puts a lower
limit on the size of the fundamental cell, about $3000$ Mpc, which
is a cube in the cases of \cite{sok} and \cite{as}. Later on, it
was shown that this result is very sensitive to the type of the 
compactifications
of the spatial sections. For a universe with spatial sections $T^{2}\times R$,
the fundamental cell's size is about $1/10$ of the horizon, and is
compatible with the homogeneity of the CMBR \cite{rouk}. In compact
universes, the pair separation
histogram would present spikes for characteristic distances. At first
it was thought that this technique, known as the \textit{crystallographic
method}, was able to totally determine the topology of the universe
\cite{lll}. It turned out that the crystallographic method only applied
when the group that defines the manifold, contained at least a
Clifford translation, i.e. a translation which moves all the points by
the same distance \cite{llu} and \cite{gtrb}. Generalisations of the 
crystallographic method were proposed, for example in \cite{fg}.

Also in compact universes the light front of the CMBR interacts with
itself producing circles in its sky pattern \cite{css2}.

Recent results have called our attention to the possibility that methods
based on multiple images will prove not to be efficient \cite{grt}.
The reasoning is that, according to observations, the curvature is
very small, so the fundamental regions are so big that there has not
been time enough for the formation of ghost images. The result is
that for low curvature universes such as ours, only compact universes
with the smallest volumes could be detected by pattern repetitions.

The knowledge of the spectrum of the Laplacian for compact spaces
have some motivations. One of them is the decomposition of the CMBR
in right set of modes, if space is non trivial. It has been studied
numerically for compact hyperbolic space by \cite{cornish1998,inoue1998}.
And more recently \cite{lehoucq2002} for compact spherical space.

In 1948, Casimir calculated an attraction force between
parallel non charged plates in vacuum \cite{Casimir48}. This force is
universal, in the sense that it does not depend on 
the physical properties of the plates, or the electric charge. 
Its origin is of topological nature, connected to the boundary
conditions. There are interesting articles, in which the 
imposition of boundary conditions is addressed in a more rigorous and 
realistic approach, for instance \cite{plb}.
 
The
Casimir effect also occurs in compact spaces. For the present era 
where the size of the Universe is $\sim 3000$ Mpc the Casimir effect 
is irrelevant. Anyway, soon after the cosmological singularity the
co-moving distances were on the order of the Planck scale. 
It is general believed that the very primordial
Universe was dominated by strict quantum effects including Casimir
effect. Although today quantum effects are commonly disregarded, 
they must have been fundamentally important in the primordial
Universe. We shall use the massless scalar field as a toy model.

The Casimir effect can be obtained by
analytic continuation of the zeta function. This procedure is known as
the zeta function regularisation, see for example \cite{bordag2001}, 
\cite{elizalde.et.al.}. Anyway, in a more realistic context, the
ability of the zeta function method has been questioned \cite{npb}.
For a calculation of the Casimir energy in the 
spherical space, see \cite{elizalde2004}.

In this work we expose a different technique which was previously
used by us for the hyperbolic space \cite{muller2001,muller2002a,muller2002b}. 

Previous calculations for the Casimir effect for the Euclidean space 
were done, for instance by, DeWitt, Hart, Isham and others \cite{dWH}. 
Particularly interesting is Dowker's article in which the result is 
also written in terms of Epstein's zeta functions. A topological 
classification of possible ``gauge fields'' over compact base manifolds, is also
given in \cite{dWH}. These and other very interesting topological
questions are not going to be addressed in this present work.                

The crystallographic groups are discrete subgroups of the full isometry 
group of the
Euclidean space. Among all crystallographic groups, only a few of them are of 
interest in connection to manifolds. These groups must be torsion-free, for 
the Euclidean and Hyperbolic spaces. In the celebrated list of problems 
proposed by Hilbert, the $18th$ was answered 
affirmatively by L. Bieberbach in 1910-1912 \cite{Wolf}, \cite{Ratcliffe}, 
see also \cite {gomero-reboucas}. 
Bieberbach showed that there are only a finite number of crystallographic 
groups for an Euclidean space of fixed dimension. For example, when
the dimension of the Euclidean space is $3$ there are only $10$
compact manifolds, of which $6$ are orientable and $4$ non orientable. 

In this present work we shall investigate the $6$ compact orientable manifolds 
and obtain the Casimir energy for each of them. 
Due to the more simple 
structure connected to these topological spaces compared to the hyperbolic or 
spherical case, an analytical result is obtained in terms of Epstein zeta 
functions. It is explicitly obtained that our result is independent of the 
type of coupling with curvature $\xi$ for these particular manifolds, which 
means that the calculated Casimir energy density is the same even in an 
inflationary de Sitter regime. 

The paper is organised as follows. Section \ref{II} is included for
completeness. In this section, the  expectation value for the energy 
momentum tensor, for a massless scalar 
field in closed Minkowski spaces is obtained. Section \ref{III} presents a
brief description of the compact flat manifolds. The Casimir energy is
obtained in each subsection, \ref{sec1}-\ref{sec6} and a numerical plot
is given. Natural units are
used, $c=G=\hbar=1$, and the metric $\eta_{\mu\nu}=[-1,1,1,1]$
\section{Casimir effect in closed spaces \label{II}}
The point splitting method was constructed
to obtain the renormalised (finite) expectation values for the quantum
mechanical operators. It is based on the Schwinger formalism \cite{Schwinger},
and was developed in the context of curved space by DeWitt \cite{dWitt}.
Further details are contained in the articles by Christensen \cite{chris1},
\cite{chris2}. For a review, see \cite{grib1994}.

Metric variations in the scalar action \[
S=-\frac{1}{2}\int\sqrt{-g}(\phi_{,\rho}\phi^{,\rho}+\xi R\phi^{2}+m^{2}\phi^{2})d^{4}x\,\,,
\]
give the classical energy-momentum
tensor \begin{eqnarray}
T_{\mu\nu} & = & (1-2\xi)\phi_{,\mu}\phi_{,\nu}
-\left( \frac{1}{2}-2\xi\right)\phi_{,\rho}\phi^{,\rho}g_{\mu\nu}
-2\xi\phi\phi_{;\mu\nu}\nonumber \\
 &  & +2\xi g_{\mu\nu}\phi\square\phi+\xi G_{\mu\nu}\phi^{2}-\frac{1}{2}m^{2}g_{\mu\nu}\phi^{2}\,\,,\label{tmunu}\end{eqnarray}
 where $G_{\mu\nu}$ is the Einstein tensor. As expected, for conformal 
coupling, $\xi=1/6$ with massless
fields, it can be verified that the trace of the above tensor is identically
zero if $m=0.$ In the following we shall restrict to spatial Euclidean 
sections and massless fields.

The renormalised energy-momentum tensor involves field products at
the same space-time point. Thus the idea is to calculate the averaged
products at separate points, $x$ and $x^{\prime}$, taking the limit
$x^{\prime}\rightarrow x$ in the end. 
\begin{equation}
\langle 0|T_{\mu\nu}\left(x\right)|0\rangle=
\frac{1}{2}\lim_{x^{\prime}\rightarrow x}T(x,x^\prime)_{\mu\nu} \label{Tnl}
\end{equation}
where for the particular manifolds addressed in this article,
\begin{eqnarray}
&&T(x,x^\prime)_{\mu\nu}=\\
&&(1-2\xi)(\left\{ 
\partial_\mu\partial_{\nu^\prime}+\partial_{\mu^\prime}\partial_\nu) -
2\xi
(
\partial_{\beta^\prime}\partial_{\alpha^\prime}+\partial_\alpha\partial_\beta)
\right\}G^{(1)}_{\mathcal{M}}(x,x^\prime) \label{Tem_recobrimento}
\end{eqnarray}
where $G^{(1)}_{\mathcal{M}}$ is the Hadamard function, which is
the expectation value of the anti-commutator of 
$\phi(x)$ and $\phi(x^{\prime})$ on the manifold $\mathcal{M}$, see below. 

The causal Green function or Feynman propagator, for the infinite 
covering space $R^3$, is obtained as \[
G(x,x^{\prime})=i\langle0|T\phi(x)\phi(x^{\prime})|0\rangle\,\,,\]
 where $T$ is the time-ordering operator. Taking its real and imaginary
parts, \begin{equation}
G(x,x^{\prime})=G_{s}(x,x^{\prime})+\frac{i}{2}G^{(1)}(x,x^{\prime})\,\,,
\label{dfg}\end{equation}
 we get for the Hadamard function \[
G^{(1)}(x,x^{\prime})=\langle0|\{\phi(x),\phi(x^{\prime})\}|0\rangle\
=2\mathop{\textrm{Im}}G(x,x^{\prime})\,\,.\]

For a massless scalar field 
\[
G^{(1)}(x,x^\prime)=\frac{1}{2\pi^2}\frac{1}{-(t-t^\prime)^2+r^2}=
\frac{1}{2\pi^2}\frac{1}{(x-x^\prime)^2}. 
\]

Green functions, as any other function defined in the spatially compact
space-time $R\times\mathcal{M}$, must have the same periodicities
of the manifold $\mathcal{M}$ itself. One way of imposing this periodicity
is by the direct summation of all the elements of the group $\Gamma$ 
which defines the manifold
\begin{equation}
G^{(1)}_{\mathcal{M}}(x,x^\prime)=\frac{1}{2\pi^2}\sum_{\gamma_i \in \Gamma}\frac{1}{-(t-t^\prime)^2+(\gamma_i\vec{r}-\vec{r}^\prime)^2}\,\,.
\label{f.autom}
\end{equation}
The above expression is named the Poincar\'{e} series, and when
it converges, it defines functions on the manifold
$\mathcal{M}$, in this particular case, the Hadamard function. 

There are recent articles were the physical 
context connected to particular boundary conditions, Dirichlet,
Neumann and periodic is addressed \cite{sugs.}. Edery, develops a
very interesting cut-off technique and obtain analytical
formulas as sums over gamma and Riemann zeta functions plus an exact
multidimensional remainder expressed as sums over modified Bessel
functions.

Anyway if the spatial sections are compact, the point $x$ and the
point $\gamma x$ where $\gamma \in \Gamma$, $x \equiv \gamma x$ for all
the elements of the fundamental group $\Gamma$. That is, the point $x$
and $\gamma x$ are the same and identical point. This means that the
only possibility is that the function (\ref{f.autom}) must be periodic 
in some sense. Mathematically, the  functions defined on a closed 
manifold $\mathcal{M}$ are called automorfic, see for example 
\cite{balazsvoros}.

In \cite{sugs.}, for instance,  
the summations are over the spectrum of the Laplace operator. It is
well known that summations over the spectrum are equivalent to
summation over the closed geodesics, also known as the method of
images, \cite{balazsvoros}. This result lies at the 
hart of the
Selberg formalism for the calculations of functional traces and is 
valid at least for manifolds, Lie groups included \cite{Camporesi}, 
see also \cite{elizalde.et.al.} 

Wald's axioms for the expectation value of the energy momentum 
tensor are satisfied \cite{axwald}. Albeit this does not prove that
the chosen state is the correct one, it does indeed indicates a
criterion for a candidate physical state. 

The Casimir energy density is independent of $\xi$ and is given by 
the $00$ component of (\ref{Tem_recobrimento}) which is the projection 
to static observers 
$\langle 0|T_{\mu\nu}\left(x\right)|0\rangle u^\mu u^\nu$ with $4$ velocity 
$u=(1,0,0,0)$
\begin{equation}
\rho^c_{\mathcal{M}}=-\frac{1}{\pi^2}\lim_{\vec{r}\rightarrow \vec{r}^\prime}
\sum_{\gamma \in \Gamma}{}^\prime\frac{1}{(\gamma\vec{r}-\vec{r}^\prime)^4},
\label{calc_e_casimir}
\end{equation}
where the $\prime$ means that the direct path with $\gamma=1$ is to be 
excluded. It can be shown that the exclusion of the direct path corresponds 
to a renormalisation of the cosmological constant \cite{muller2002a}.
\section{The closed oriented Euclidean spaces \label{III}}
Differently from the hyperbolic or spherical spaces, only a finite number of 
topologies can be modelled in the Euclidean space. A representation for an 
arbitrary element for the rigid motion subgroup of the full 
isometry group can be given 
\begin{equation} 
\gamma=\left(\begin{array}{cccc}
 & & &a_1\\
 &R & &a_2\\
 & & &a_3\\
0 &0 &0 &1\\
\end{array}
\right),
\label{representacao}
\end{equation}
where $R \in SO(3)$ is an orthogonal rotation matrix and $\vec{a}$ is 
the translation. 
The action of the group on any point $x=\{x_1,x_2,x_3\}$ of $E^3$ is linear and 
multiplicative 
\[
\gamma x=\left(\begin{array}{cccc}
 & & &a_1\\
 &R & &a_2\\
 & & &a_3\\
0 &0 &0 &1\\
\end{array}
\right) 
\left( 
\begin{array}{c}
x_1\\
x_2\\
x_3\\
1\\
\end{array}
\right)
\]

The discrete subgroups of the isometry group define the manifolds modelled 
in the space of interest.
A group $\Gamma$ is said to be torsion-free if for each element 
$\gamma\in \Gamma$, $\gamma^n\neq 1$ for any integer $n$. In the Euclidean 
case, the torsion-free 
elements must be ``screw motions", i.e. a combined rotation and translation 
along the rotation 
axis. The holonomy group is the rotational part of $\Gamma$, $\Psi=\Gamma/T$, 
where $T$ is the translational part of the group.  
Two oriented manifolds are diffeomorfic if and only if the discrete 
torsion-free groups have isomorfic holonomy groups, $\Psi$. Of course, 
$\Psi\subset SO(3)$ is a finite group. According to Lemma 3.5.4 of Wolf's book 
\cite{Wolf}, for the $E^3$ case, the only possibilities for $\Psi$ are 
$\Psi=\mathbb{Z}_1,\mathbb{Z}_2,\mathbb{Z}_4,
\mathbb{Z}_3,\mathbb{Z}_6,\mathbb{Z}_2\otimes \mathbb{Z}_2$, which
give rise to to the $6$ closed oriented manifolds \cite{Wolf}. These 
manifolds and the Casimir energy are given in the subsections 
\ref{sec1}-\ref{sec6}.

The very important Theorem 8.2.5 of Ratcliffe's book \cite{Ratcliffe} is 
reproduced in the following \\

{\it Every compact, $n$-dimensional, Euclidean space-form is finitely 
covered by an Euclidean n-torus.} \\ 

This theorem is in fact Bieberbach's first theorem, which states that 
$T$ is a normal subgroup of finite index in $\Gamma$. This finite index 
is given by the order of $\Psi$ and corresponds to the number of coverings of 
the space in question by the Euclidean $n$-torus. The above theorem is
used to rewrite the summations in 
the expectation values over all the elements of the group as a finite sum 
of Epstein type zeta functions. The Epstein zeta function is already well 
known to give the appropriate result for the Casimir energy for the $n-$ 
torus, see for example \cite{bv} \cite{elizalde.et.al.}. In the original 
Epstein's notation let $p$ be a positive integer and 
\begin{eqnarray*}
&&\vec{g}=(g_1,...,g_p), \;\; g_i \in \mbox{R}, \;\; \vec{g} \in\mbox{R}^p\\
&&\vec{h}=(h_1,...,h_p),\;\; h_i\in \mbox{R},\;\; \vec{h}\in\mbox{R}^p\\
&&\vec{m}=(m_1,...,m_p),\;\; m_i\in \mbox{Z},\;\; \vec{m}\in \mbox{Z}^p. 
\end{eqnarray*}
The scalar product is defined as 
\[
(\vec{g},\vec{h})=\sum_{\nu=1}^p g_\nu h_\nu,
\]
and $c_{\mu\nu}$ an invertible $p\times p$ matrix associated with a quadratic 
form as follows 
\[
\varphi(x)=\sum_{\mu,\nu=1}^{p}c_{\mu\nu}x_\mu x_\nu.
\]
With the above definitions Epstein's zeta function is written as 
\begin{equation}
Z\left|\begin{array}{ccc} g_1 &....&g_p\\ h_1 & ....& h_p\end{array}
\right|(s)_{\varphi}=\sum_{m_1,...,m_p=-\infty}^\infty{}^\prime[\varphi(\vec{m}+\vec{g})]^
{-\frac{s}{2}}e^{2\pi i(\vec{m},\vec{h})},
\label{fze}
\end{equation}
where the $\prime$ means that $m_1=0,\; m_2=0,\;...m_p=0$ is to be excluded.
Generalisations for non homogeneous Epstein zeta functions and reflection 
formulae, can be found in \cite{e.1998}, \cite{elizalde.et.al.}.
\subsection{$\Psi=\mathbb{Z}_1$ \label{sec1} }
\begin{figure}[tpb]
\includegraphics[scale=0.4]{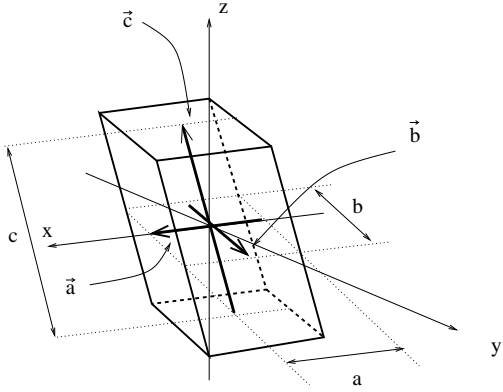}
\caption{Dirichlet Region}
\label{fig1}
\end{figure}

The volume of this manifold is $V=ab_2c_3$. The Dirichlet region is the 
parallelepiped given in Figure \ref{fig1}
According to the representation (\ref{representacao}), the 
matrices 
\begin{eqnarray}
&&\gamma_1=\left(\begin{array}{cccc}
1 &0 &0 &a\\
0 &1 &0 &0\\
0 &0 &1 &0\\
0 &0 &0 &1\\
\end{array} \right),
\;\;\gamma_2=\left(\begin{array}{cccc}
1 &0 &0 &b_1\\
0 &1 &0 &b_2\\
0 &0 &1 &0\\
0 &0 &0 &1\\
\end{array} \right),\nonumber\\
&&\gamma_3=\left(\begin{array}{cccc}
1 &0 &0 &c_1\\
0 &1 &0 &c_2\\
0 &0 &1 &c_3\\
0 &0 &0 &1\\
\end{array} \right)
\end{eqnarray}
with the respective inverses are also the generators of the translational 
part of the group $\Gamma=T=\{\gamma_1,\gamma_2,\gamma_3\}$. 
The identifications are as follows 
\[
\vec{r}=(x,y,z)\rightarrow (x+ma+nb_1+kc_1,y+nb_2+kc_2,z+kc_3)
\]
and according to (\ref{calc_e_casimir})
\begin{eqnarray*}
&&\rho^c_{\mathbb{Z}_1}=-\frac{1}{\pi^2}\sum_{m,n,k=-\infty}^{+\infty}{}^\prime\left\{
\left[(ma+nb_1+kc_1)^2 \right.\right. \\
&&\left. \left. +(nb_2+kc_2)^2+(kc_3)^2\right]^{-2}\right\}
\end{eqnarray*}
\begin{equation}
\rho^c_{\mathbb{Z}_1}=-\frac{1}{\pi^2}
Z\left|\begin{array}{ccc} 0 & 0 &0\\ 0 & 0 & 0 \end{array}\right|(4)_\varphi,
\end{equation}
where according to (\ref{fze}) the matrix 
\[
c_{\mu\nu}=\left(
\begin{array}{ccc} 
a^2 & ab_1 & ac_1\\ 
ab_1 &b_1^2+b_2^2 & b_1c_1+b_2c_2 \\
ac_1 & b_1c_1+b_2c_2 & c_1^2+c_2^2+c_3^2 
\end{array}\right)
\]
defines the positive quadratic form $\varphi
(m)=c_{11}m^2+c_{22}n^2+c_{33}k^2
+2c_{12}mn+2c_{13}mk+2c_{23}nk$.
\subsection{$\Psi=\mathbb{Z}_2$ \label{sec2} } 
\begin{figure}[tpb]
\includegraphics[scale=0.4]{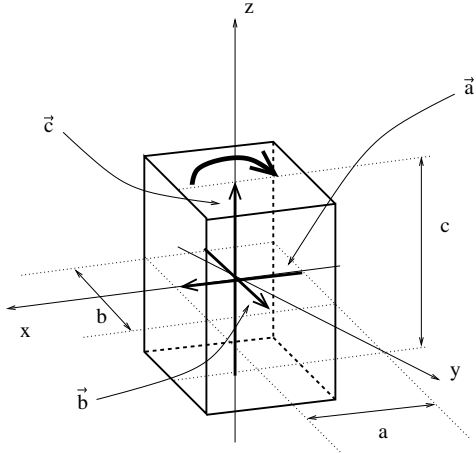}
\caption{Dirichlet Region, DR. The opposite faces in the $z$ direction are 
identified with a rotation of $\pi$.}
\label{fig2}
\end{figure}

The volume of this manifold is $V=ab_2c$ Figure \ref{fig2}. 
Now the matrices and their inverses 
\begin{eqnarray}
&&\gamma_1=\left(\begin{array}{cccc}
1 &0 &0 &a\\
0 &1 &0 &0\\
0 &0 &1 &0\\
0 &0 &0 &1\\
\end{array} \right),
\;\;\gamma_2=\left(\begin{array}{cccc}
1 &0 &0 &b_1\\
0 &1 &0 &b_2\\
0 &0 &1 &0\\
0 &0 &0 &1\\
\end{array} \right),\nonumber\\
&&\gamma_3=\left(\begin{array}{cccc}
-1 &0 &0 &0\\
0 &-1 &0 &0\\
0 &0 &1 &c\\
0 &0 &0 &1\\
\end{array} \right)
\end{eqnarray}
generate the group $\Gamma$.

The generators of the translational part $T$, now, are given by 
$T=\Gamma/\mathbb{Z}_2=\{\gamma_1,\gamma_2,\gamma_3^2\}$, together with the
inverses. The fundamental 
domain, $FD$, for $\Gamma/\mathbb{Z}_2$ has volume $V=2ab_2c$ and is given by 
two copies of the Dirichlet region of Figure \ref{fig2};  
$FD=\{1.DR,\gamma_3.DR\}$. 

The identifications in this manifold are provided by 
\begin{eqnarray*}
&&\vec{r}=(x,y,z)\rightarrow (x+ma+nb_1,y+nb_2,z+2kc) \\
&&\rightarrow (-x+ma+nb_1,-y+nb_2,z+(2k+1)c)
\end{eqnarray*}

\begin{figure}[tpb]
\includegraphics[scale=0.4]{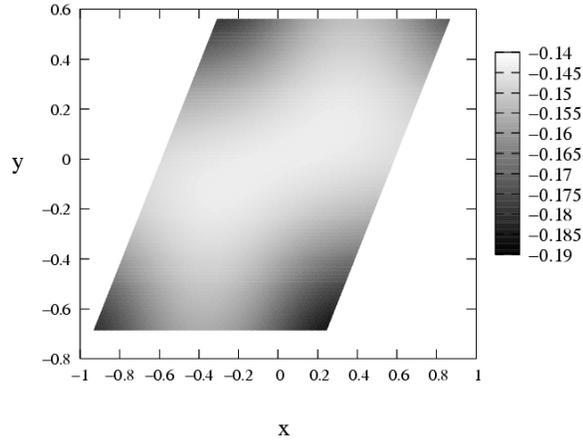}
\caption{The Casimir energy density in natural units, eq. (\ref{Ec1})
inside the DR. in Figure \ref{fig2}. The $z$ coordinate is chosen to be $z=0.6$}
\label{grf1}
\end{figure}

Using (\ref{calc_e_casimir}), the Casimir energy is given by 
\begin{eqnarray*}
&&\rho^c_{\mathbb{Z}_2}=-\frac{1}{\pi^2}\sum_{m,n,k=-\infty}^{+\infty}{^\prime}\left\{
\left[(ma+nb_1)^2 +(nb_2)^2+(2kc)^2\right]^{-2}\right.\\ 
&&\left.+\left[(-2x+ma+nb_1)^2 +
(-2y+nb_2)^2 +((2k+1)c)^2\right]^{-2}\right\}
\end{eqnarray*}
\begin{equation}
\rho^c_{\mathbb{Z}_2}=-\frac{1}{\pi^2}\left(
Z\left|\begin{array}{ccc} 0 & 0 &0\\ g_x & g_y & g_z \end{array}\right|(4)_\varphi+
Z\left|\begin{array}{ccc} 0 & 0 &0\\ 0 & 0 & 0 \end{array}\right|(4)_\varphi \right)
\label{Ec1}
\end{equation}
where 
\begin{eqnarray}
&&\vec{g}=\left(-\frac{2x}{a}+\frac{2yb_1}{ab_2},-\frac{2y}{b_2},1/2 \right) \\
&&c_{\mu\nu}=\left(
\begin{array}{ccc} 
a^2 & ab_1 & 0\\ 
ab_1 &b_1^2+b_2^2 & 0 \\
0 & 0 & 4c^2 
\end{array}\right)
\end{eqnarray}
The values of $a=1.2$, $b_1=0.7$, $b_2=1.4$ and $c=1.3$ are chosen. 
The truncation in (\ref{fze}) is done such that the summation is
over a geodesic sphere of radius $R\sim 100$ of integer numbers. The 
corresponding truncation error on the energy density should be less
than $1\%$. The result is shown in Figure \ref{grf1}, and is independent 
of the coordinate $z$.
\subsection{$\Psi=\mathbb{Z}_4$ \label{sec3}}
\begin{figure}[tpb]
\includegraphics[scale=0.4]{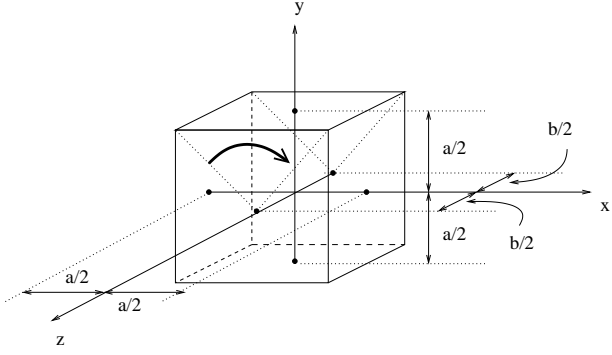}
\caption{Dirichlet Region. The opposite faces in the $z$ direction are 
identified with a rotation of $\pi/2$}
\label{fig3}
\end{figure}

\begin{figure}[tpb]
\includegraphics[scale=0.4]{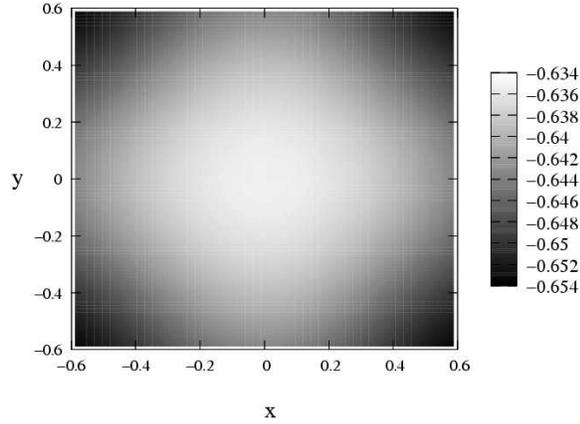}
\caption{The Casimir energy density in natural units, given in (\ref{Ec2}) 
inside the DR.  in Figure \ref{fig3}, with the coordinate $z=0.6$}
\label{grf2}
\end{figure}

The volume of this manifold is $V=a^2b$ Figure \ref{fig3}. 
The matrices
\begin{eqnarray}
&&\gamma_1=\left(\begin{array}{cccc}
1 &0 &0 &a\\
0 &1 &0 &0\\
0 &0 &1 &0\\
0 &0 &0 &1\\
\end{array} \right),
\;\;\gamma_2=\left(\begin{array}{cccc}
1 &0 &0 &0\\
0 &1 &0 &a\\
0 &0 &1 &0\\
0 &0 &0 &1\\
\end{array} \right),\nonumber\\
&&\gamma_3=\left(\begin{array}{cccc}
0 &-1 &0 &0\\
1 &0 &0 &0\\
0 &0 &1 &b\\
0 &0 &0 &1\\
\end{array} \right),
\end{eqnarray}
together with their inverses are the generators of $\Gamma$. The 
translational part is generated by 
$T=\Gamma/\mathbb{Z}_4=\{\gamma_1,\gamma_2,\gamma_3^{4}\}$, together
with the inverses. The fundamental 
region, $FR$ for $\Gamma/\mathbb{Z}_4$ has volume $V=4a^2b$ and is given by $4$ copies of the DR in 
Figure \ref{fig3}; $FR=\{DR, \gamma_3.DR,\gamma_3^{2}.DR,\gamma_3^{3}.DR\}$. The 
identifications and the direct summation are given in the  \ref{ap1}.
The result for the Casimir energy according to (\ref{calc_e_casimir}) and (\ref{fze}) is the following
\begin{eqnarray}
&&\rho^c_{\mathbb{Z}_4}=-\frac{1}{\pi^2}\left(
Z\left|\begin{array}{ccc} 0 & 0 &0\\ 0 & 0 & 0 \end{array}\right|(4)_\varphi+
Z\left|\begin{array}{ccc} 0 & 0 &0\\
      g^1_x & g^1_y & g^1_z \end{array}\right|(4)_\varphi\right.\nonumber\\
&&\left.+Z\left|\begin{array}{ccc} 0 & 0 &0\\
      g^2_x & g^2_y & g^2_z \end{array}\right|(4)_\varphi+ 
 Z\left|\begin{array}{ccc} 0 & 0 &0\\
      g^3_x & g^3_y & g^3_z \end{array}\right|(4)_\varphi  \right)   
\label{Ec2}
\end{eqnarray}
where 
\begin{eqnarray}
&&\vec{g}^1=\left(-\frac{x+y}{a},\frac{x-y}{a},1/4 \right) \nonumber\\
&&\vec{g}^2=\left(\frac{-2x}{a},\frac{-2y}{a},2/4 \right) \nonumber\\
&&\vec{g}^3=\left(\frac{y-x}{a},-\frac{x+y}{a},3/4 \right) \nonumber\\
&&c_{\mu\nu}=\left(
\begin{array}{ccc} 
a^2 & 0 & 0\\ 
0 &a^2 & 0 \\
0 & 0 & 16b^2 
\end{array}\right)
\end{eqnarray}
The values of $a=1.2$, $b=1.4$ are chosen. 
The truncation in (\ref{fze}) is done such that the summation is
over a geodesic sphere of radius $R\sim 100$ of integer numbers. The 
corresponding truncation error on the energy density should be less
than $1\%$. The result is shown in Figure \ref{grf2}, and is independent
of the coordinate $z$
\subsection{$\Psi=\mathbb{Z}_3$ \label{sec4}}
\begin{figure}[tpb]
\includegraphics[scale=0.4]{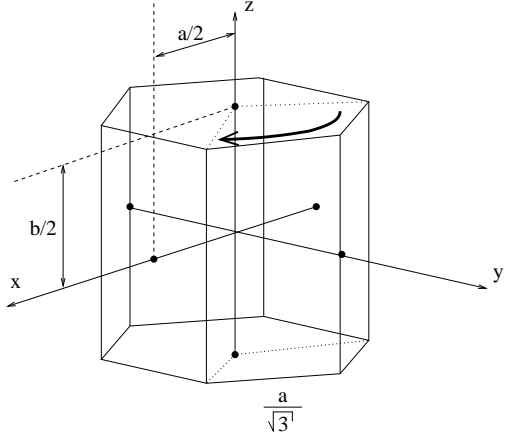}
\caption{Dirichlet Region. The top and bottom faces in the $z$ direction are 
identified with a rotation of $2\pi/3$}
\label{fig4}
\end{figure}

The volume of this manifold is $V=\sqrt{3}a^2b$. The Dirichlet domain is an 
hexagonal prism of side $a/\sqrt{3}$ Figure \ref{fig4}
The face pairing matrices 
\begin{eqnarray}
&&\gamma_1=\left(\begin{array}{cccc}
1 &0 &0 &a\\
0 &1 &0 &0\\
0 &0 &1 &0\\
0 &0 &0 &1\\
\end{array} \right),
\;\;\gamma_2=\left(\begin{array}{cccc}
1 &0 &0 &-a/2\\
0 &1 &0 &\sqrt{3}a/2\\
0 &0 &1 &0\\
0 &0 &0 &1\\
\end{array} \right),\nonumber\\
&&\gamma_3=\left(\begin{array}{cccc}
1 &0 &0 &a/2\\
0 &1 &0 &\sqrt{3}a/2\\
0 &0 &1 &0\\
0 &0 &0 &1\\
\end{array} \right),\nonumber\\
&&\gamma_4=\left(\begin{array}{cccc}
-1/2 &-\sqrt{3}/2 &0 &0\\
\sqrt{3}/2 &-1/2 &0 &0\\
0 &0 &1 &b\\
0 &0 &0 &1\\
\end{array} \right).
\end{eqnarray}
together with the inverses generate the group $\Gamma$. 

\begin{figure}[tpb]
\includegraphics[scale=0.4]{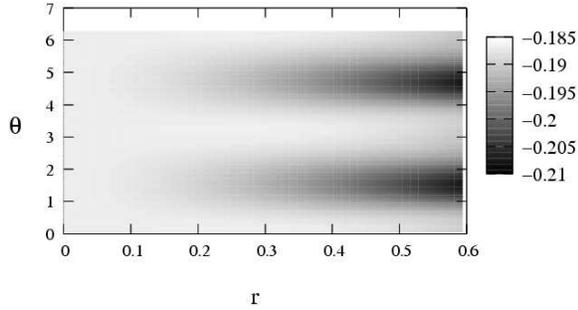}
\caption{The Casimir energy density in natural units, 
given in (\ref{Ec3}) inside the DR. in Figure \ref{fig4} with the 
coordinate $z=0.6$. The region is a disc of radius $r=0.6$, $\theta$ 
is the angle between the $x$ and $y$ coordinates: $\theta=0$ corresponds to the
$x$ axis of Figure \ref{fig4}.}
\label{grf3}
\end{figure}

The generators of the  translational 
part are given by 
$T=\Gamma/\mathbb{Z}_3=\{\gamma_1,\gamma_3,\gamma_4^3\}$, with the inverses. The fundamental 
region, $FR$ for $\Gamma/\mathbb{Z}_3$ has volume $V=3\sqrt{3}a^2b$ and is given by $3$ 
copies of the DR in 
Figure \ref{fig4}; $FR=\{DR, \gamma_4.DR,\gamma_4^{2}.DR\}$. The 
identifications are given in the \ref{ap2}. The Casimir energy, 
according to (\ref{calc_e_casimir}) and (\ref{fze}) is
\begin{eqnarray}
&&\rho^c_{\mathbb{Z}_3}=-\frac{1}{\pi^2}\left(
Z\left|\begin{array}{ccc} 0 & 0 &0\\ 0 & 0 & 0 \end{array}\right|(4)_\varphi+
Z\left|\begin{array}{ccc} 0 & 0 &0\\
      g^1_x & g^1_y & g^1_z \end{array}\right|(4)_\varphi\right.\nonumber\\
&&\left.+Z\left|\begin{array}{ccc} 0 & 0 &0\\
      g^2_x & g^2_y & g^2_z \end{array}\right|(4)_\varphi \right)   
\label{Ec3}
\end{eqnarray}
where 
\begin{eqnarray}
&&\vec{g}^1=\left(\frac{-2x}{a},\frac{x-y\sqrt{3}}{a},1/3 \right) \nonumber\\
&&\vec{g}^2=\left(-\frac{x-y\sqrt{3}}{a},-\frac{x+y\sqrt{3}}{a},2/3 \right) \nonumber\\
&&c_{\mu\nu}=\left(
\begin{array}{ccc} 
a^2 & a^2/2 & 0\\ 
a^2/2 &a^2 & 0 \\
0 & 0 & 9b^2 
\end{array}\right)
\end{eqnarray}
The values of $a=1.2$, $b=1.4$ are chosen. 
The truncation in (\ref{fze}) is done such that the summation is
over a geodesic sphere of radius $R\sim 100$ of integer numbers. The 
corresponding truncation error on the energy density should be less
than $1\%$. The result is shown in Figure \ref{grf3}, and is
independent of the $z$ coordinate.
\subsection{$\Psi=\mathbb{Z}_6$ \label{sec5}}
\begin{figure}[tpb]
\includegraphics[scale=0.4]{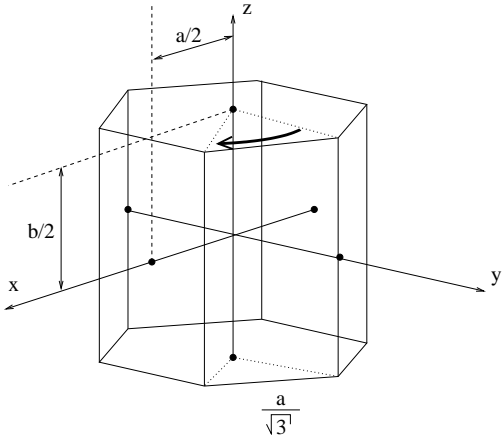}
\caption{Dirichlet Region. The top and bottom faces in the $z$ direction are 
identified with a rotation of $\pi/3$}
\label{fig5}
\end{figure}

The Dirichlet domain is a prism with hexagonal base of side $a/\sqrt{3}$, 
Figure \ref{fig5}
The volume of this manifold is $V=\sqrt{3}a^2b$. The face pairing matrices are 
\begin{eqnarray}
&&\gamma_1=\left(\begin{array}{cccc}
1 &0 &0 &a\\
0 &1 &0 &0\\
0 &0 &1 &0\\
0 &0 &0 &1\\
\end{array} \right),
\;\;\gamma_2=\left(\begin{array}{cccc}
1 &0 &0 &-a/2\\
0 &1 &0 &\sqrt{3}a/2\\
0 &0 &1 &0\\
0 &0 &0 &1\\
\end{array} \right),\nonumber\\
&&\gamma_3=\left(\begin{array}{cccc}
1 &0 &0 &a/2\\
0 &1 &0 &\sqrt{3}a/2\\
0 &0 &1 &0\\
0 &0 &0 &1\\
\end{array} \right),\nonumber\\
&&\gamma_4=\left(\begin{array}{cccc}
1/2 &-\sqrt{3}/2 &0 &0\\
\sqrt{3}/2 &1/2 &0 &0\\
0 &0 &1 &b\\
0 &0 &0 &1\\
\end{array} \right),
\end{eqnarray}
which together with their inverses generate $\Gamma$. The translational part 
is generated by  
$T=\Gamma/\mathbb{Z}_6=\{\gamma_1,\gamma_3,\gamma_4^6 \}$, 
together with the inverses. The fundamental 
region, $FR$ for $\Gamma/\mathbb{Z}_6$ has volume $V=6\sqrt{3}a^2b$ and is 
given by $6$ 
copies of the DR in Figure \ref{fig5}; $FR=\{DR, \gamma_4.DR,\gamma_4^{2}.DR,
\gamma_4^{3}.DR,\gamma_4^{4}.DR\,\gamma_4^{5}.DR\}$. The 
identifications are given in \ref{ap3}

\begin{figure}[tpb]
\includegraphics[scale=0.4]{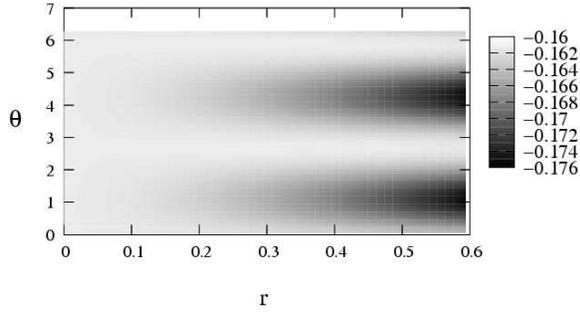}
\caption{The Casimir energy density in natural units, 
given in (\ref{Ec4}) inside the DR. in Figure 
\ref{fig5}, with the coordinate $z=0.6$. 
The region is a disc of radius $r=0.6$, $\theta$ 
is the angle between the $x$ and $y$ coordinates: $\theta=0$ corresponds to the
$x$ axis of Figure \ref{fig5}.}
\label{grf4}
\end{figure}

With a reasoning analogous to the previous sections the following result is 
obtained using (\ref{calc_e_casimir}) and (\ref{fze})
\begin{eqnarray}
&&\rho^c_{\mathbb{Z}_6}=-\frac{1}{\pi^2}\left(
Z\left|\begin{array}{ccc} 0 & 0 &0\\ 0 & 0 & 0 \end{array}\right|(4)_\varphi+
Z\left|\begin{array}{ccc} 0 & 0 &0\\
      g^1_x & g^1_y & g^1_z \end{array}\right|(4)_\varphi\right.\nonumber\\
&&\left.+Z\left|\begin{array}{ccc} 0 & 0 &0\\
      g^2_x & g^2_y & g^2_z \end{array}\right|(4)_\varphi+ 
 Z\left|\begin{array}{ccc} 0 & 0 &0\\
      g^3_x & g^3_y & g^3_z \end{array}\right|(4)_\varphi \right.\nonumber\\
 &&\left.+Z\left|\begin{array}{ccc} 0 & 0 &0\\
      g^4_x & g^4_y & g^4_z \end{array}\right|(4)_\varphi+ 
 Z\left|\begin{array}{ccc} 0 & 0 &0\\
      g^5_x & g^5_y & g^5_z \end{array}\right|(4)_\varphi \right)        
\label{Ec4}
\end{eqnarray}
where 
\begin{eqnarray}
&&\vec{g}^1=\left(-\frac{x+y/\sqrt{3}}{a},\frac{x-y/\sqrt{3}}{a},1/6 \right) \nonumber\\
&&\vec{g}^2=\left(\frac{-2x}{a},\frac{x-y\sqrt{3}}{a},2/6 \right) \nonumber\\
&&\vec{g}^3=\left(\frac{-2x+2y/\sqrt{3}}{a},\frac{-4y/\sqrt{3}}{a},3/6 \right) \nonumber\\
&&\vec{g}^4=\left(\frac{-x+y\sqrt{3}}{a},-\frac{x+y\sqrt{3}}{a},4/6 \right) \nonumber\\
&&\vec{g}^5=\left(\frac{2y/\sqrt{3}}{a},-\frac{x+y/\sqrt{3}}{a},5/6 \right) \nonumber\\
&&c_{\mu\nu}=\left(
\begin{array}{ccc} 
a^2 & a^2/2 & 0\\ 
a^2/2 &a^2 & 0 \\
0 & 0 & 36b^2 
\end{array}\right)
\end{eqnarray}
The values of $a=1.2$, $b=1.4$ are chosen. 
The truncation in (\ref{fze}) is done such that the summation is
over a geodesic sphere of radius $R\sim 100$ of integer numbers. The 
corresponding truncation error on the energy density should be less
than $1\%$. The result is shown in Figure \ref{grf4}, and is independent
of the $z$ coordinate.
\subsection{$\Psi=\mathbb{Z}_2\otimes\mathbb{Z}_2$ \label{sec6}}
\begin{figure}[tpb]
\includegraphics[scale=0.5]{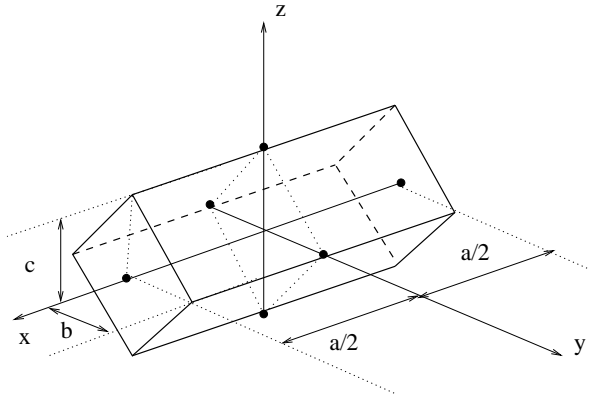}
\caption{Dirichlet Region. The opposite faces in the $x$ direction are 
identified with a rotation of $\pi$. The adjacent faces are identified with 
a rotation of $\pi$, also.}
\label{fig6}
\end{figure}

\begin{figure}[tpb]
\includegraphics[scale=0.4]{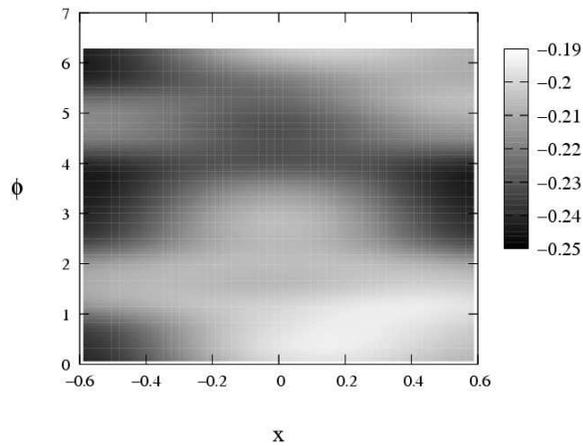}
\caption{The Casimir energy density in natural units, given in (\ref{Ec5}) 
inside the DR. in Figure \ref{fig6}, a cylindrical region surrounding the $x$
axis, of radius $0.762$, $\phi$ is the angle.}
\label{grf5}
\end{figure}

The face pairing matrices are now
\begin{eqnarray}
&&\gamma_1=\left(\begin{array}{cccc}
1 &0 &0 &a\\
0 &-1 &0 &0\\
0 &0 &-1 &0\\
0 &0 &0 &1\\
\end{array} \right),
\;\;\gamma_2=\left(\begin{array}{cccc}
-1 &0 &0 &0\\
0 &1 &0 &b\\
0 &0 &-1 &c\\
0 &0 &0 &1\\
\end{array} \right),\nonumber\\
&&\gamma_3=\left(\begin{array}{cccc}
-1 &0 &0 &0\\
0 & 1 &0 &-b\\
0 &0 &-1 &-c\\
0 &0 &0 &1\\
\end{array} \right),
\end{eqnarray}
which together with their inverses generate $\Gamma$, and the volume of the manifold is $V=2abc$, 
Figure \ref{fig6}. 
The translational part is generated by 
$T=\Gamma/(\mathbb{Z}_2\otimes\mathbb{Z}_2)=\{\gamma_1^2,\gamma_2^2,
\gamma_2\gamma_3 \}$, with the inverses. The fundamental 
region, $FR$ for $\Gamma/(\mathbb{Z}_2\otimes \mathbb{Z}_2)$ has volume 
$V=8abc$ and is 
given by $4$ 
copies of the DR in Figure \ref{fig6}; $FR=\{DR, \gamma_1.DR,\gamma_2.DR,
\gamma_1\gamma_3.DR\}$. The identifications are given in \ref{ap4}. 
With (\ref{calc_e_casimir}) and (\ref{fze}) the Casimir energy results in
\begin{eqnarray}
&&\rho^c_{\mathbb{Z}_2\otimes \mathbb{Z}_2}=-\frac{1}{\pi^2}\left(
Z\left|\begin{array}{ccc} 0 & 0 &0\\ 0 & 0 & 0 \end{array}\right|(4)_\varphi+
Z\left|\begin{array}{ccc} 0 & 0 &0\\
      g^1_x & g^1_y & g^1_z \end{array}\right|(4)_\varphi\right.\nonumber\\
&&\left.+Z\left|\begin{array}{ccc} 0 & 0 &0\\
      g^2_x & g^2_y & g^2_z \end{array}\right|(4)_\varphi+ 
 Z\left|\begin{array}{ccc} 0 & 0 &0\\
      g^3_x & g^3_y & g^3_z \end{array}\right|(4)_\varphi \right)    
\label{Ec5}
\end{eqnarray}
where 
\begin{eqnarray}
&&\vec{g}^1=\left(1/2,\frac{-y}{b},\frac{-z}{c}\right) \nonumber\\
&&\vec{g}^2=\left(\frac{-x}{a},1/2,\frac{c-2z}{2c} \right) \nonumber\\
&&\vec{g}^3=\left(\frac{a-2x}{2a},\frac{b-2y}{2b},1/2\right) \nonumber\\
&&c_{\mu\nu}=\left(
\begin{array}{ccc} 
4a^2 & 0 & 0\\ 
0 &4b^2 & 0 \\
0 & 0 & 4c^2 
\end{array}\right).
\end{eqnarray}
The values of $a=1.2$, $b=1.4$ and $c=1.3$ are chosen. 
The truncation in (\ref{fze}) is done such that the summation is
over a geodesic sphere of radius $R\sim 100$ of integer numbers. The 
corresponding truncation error on the energy density should be less
than $1\%$. The result is shown in Figure \ref{grf5}. This is the only 
manifold for which there is not an axis such that the energy density 
expectation value is constant.
\section{Conclusions}
The Casimir energy for the oriented closed Euclidean space-times are 
obtained in this work.  

Closed manifolds are described by the discrete subgroups of the
isometry group. The crystallographic groups are discrete subgroups of
the full isometry group of 
Euclidean space. Among all crystallographic groups, only a few of them are of 
interest in connection to manifolds. These groups must be torsion-free, for 
the Euclidean and Hyperbolic spaces. In the celebrated list of problems 
proposed by Hilbert, the $18th$ was answered 
affirmatively by L. Bieberbach in 1910-1912 \cite{Wolf}, \cite{Ratcliffe}, 
see also \cite{gomero-reboucas}. 
Bieberbach showed that there are only a finite number of crystallographic 
groups for an Euclidean space of fixed dimension. 

When the dimension of the Euclidean space is $3$ there are only $10$ compact 
manifolds, of which $6$ are orientable and $4$ non orientable. 

In this present work we have investigated the $6$ compact orientable manifolds 
and obtain the Casimir energy for each of them. Due to the more simple 
structure connected to these topological spaces compared to the hyperbolic or 
spherical case, an analytical result is obtained in terms of Epstein zeta 
functions. It is explicitly obtained that our result is independent of the 
type of coupling with curvature $\xi$ for these particular manifolds, 
which means that the calculated Casimir energy density is the same even in 
an inflationary de Sitter regime.

According to the theorem in Section \ref{II},  it is always possible to find
the normal subgroup of finite index, $T$. With this, the $n-$ torus
that covers the manifold a finite number of times is obtained. 
The Casimir energy is given by a finite sum of Epstein type zeta
functions. The number of distinct zeta functions is given by the
index of $T$ in $\Gamma$, which is the number of times the manifold 
is covered by this $n-$ torus. 

A numerical plot inside the Dirichlet Region for each manifold is
given. The manifold in Section \ref{sec1} is the only one for which the
Casimir energy density is homogeneous. For the manifolds 
Sections \ref{sec2}- \ref{sec5} the Casimir energy density depends on the 
$x$ and $y$ coordinates inside the fundamental domain.
For the last manifold in Section
\ref{sec6}, the Casimir energy density 
depends in all the coordinates. 

The sign of Casimir energy density is 
negative for all manifolds. Remind that there is a subtraction of the 
divergence of the infinite $R^3$ covering space, which means that the 
energy of the manifold is less than the energy of the infinite $R^3$. 
The effect of choosing different representations for the fundamental group 
$\Gamma$ acting on the field, like twisting, can increase the energy of the 
vacuum \cite{dWH}. deWitt finds that the sign of the energy depends 
on the representation of the fundamental group, the orientability of the 
manifold and the spin of the field. This must have a more fundamental reason 
connected to the Atiyah-Singer theorem, and is beyond our knowledge.
 
It is explicitly checked that the truncation error should be on the order 
of $1/R$ where $R$ is the geodesic radius of the sphere of integer numbers 
where the truncation of (\ref{calc_e_casimir}) was performed. It was used 
$R\sim 100$ such that the truncation error should be on the order of 
$1\%$. The respective values of the parameters of the manifold are
indicated in each case.
\ack
M. P. L. wishes to thank CNPq for financial support. D. M. wishes to
thank the Brazilian project {\it Nova F\'\i sica no Espa\c co} of
FAPESP.    
\appendix 
\section{}
\subsection{$\Psi=\mathbb{Z}_4$\label{ap1}}
\begin{eqnarray*}
&&\vec{r}=(x,y,z)\rightarrow (x+ma,y+na,z+4kb) \\
&&\rightarrow (-y+ma,x+na,z+(4k+1)b)\\
&&\rightarrow (-x+ma,-y+na,z+(4k+2)b)\\
&&\rightarrow (y+ma,-x+na,z+(4k+3)b)
\end{eqnarray*}
According to the above the Casimir energy is given by 
\begin{eqnarray*}
&&\rho^c_{\mathbb{Z}_4}=-\frac{1}{\pi^2}\sum_{m,n,k=-\infty}^{+\infty}{}^\prime\left\{
\left[(ma)^2 +(na)^2+(4kb)^2\right]^{-2}\right.\\ 
&&\left.  +\left[(-x-y+ma)^2 +(x-y+na)^2 +((4k+1)b)^2\right]^{-2}\right.\\
&&\left.+\left[(-2x+ma)^2 +(-2y+na)^2 +((4k+2)b)^2\right]^{-2} \right.\\
&&\left. +\left[(y-x+ma)^2 +(-x-y+na)^2 +((4k+3)b)^2\right]^{-2} \right\}
\end{eqnarray*}
\subsection{$\Psi=\mathbb{Z}_3$\label{ap2}}
\begin{eqnarray*}
&&\vec{r}=(x,y,z)\rightarrow [x+ma+na/2,y+na\sqrt{3}/2,z+3kb] \\
&&\rightarrow [-(x+y\sqrt{3})/2+ma+na/2, \\ 
&& (x\sqrt{3}-y)/2+na\sqrt{3}/2,z+(3k+1)b]\\
&&\rightarrow [-(x-y\sqrt{3})/2+ma+na/2, \\ 
&&-(x\sqrt{3}+y)/2+na\sqrt{3}/2,z+(3k+2)b]
\end{eqnarray*}
According to the above the Casimir energy is given by 
\begin{eqnarray*}
&&\rho^c_{\mathbb{Z}_3}=-\frac{1}{\pi^2}\sum_{m,n,k=-\infty}^{+\infty}{}^
\prime\left\{
\left[(ma+na/2)^2 +(na\sqrt{3}/2)^2\right. \right.\\ 
&&\left. \left.+(3kb)^2\right]^{-2}+\left
     [(-(3x+y\sqrt{3})/2+ma+na/2)^2 
\right. \right. \\
&&\left. \left.+
( (\sqrt{3}x-3y)/2+na\sqrt{3}/2 )^2 
+((3k+1)b)^2\right]^{-2}\right.\\
&&\left.+\left[(-(3x-y\sqrt{3})/2+ma+na/2)^2 \right.\right. \\
&&+\left.\left.(-(\sqrt{3}x+3y)/2+na\sqrt{3}/2)^2 +((3k+2)b)^2\right]^{-2} 
\right\}
\end{eqnarray*}
\subsection{$\Psi=\mathbb{Z}_6$\label{ap3}}
\begin{eqnarray*}
&&\vec{r}=(x,y,z)\rightarrow [x+ma+na/2,y+na\sqrt{3}/2,z+6kb] \\
&&\rightarrow [(x-y\sqrt{3})/2+ma+na/2,\\
&&(\sqrt{3}x+y)/2+na\sqrt{3}/2,z+(6k+1)b]\\
&&\rightarrow [-(x+y\sqrt{3})/2+ma+na/2, \\ 
&& (x\sqrt{3}-y)/2+na\sqrt{3}/2,z+(6k+2)b]\\
&&[-x+ma+na/2,-y+na\sqrt{3}/2,z+(6k+3)b]\\
&&\rightarrow [-(x-y\sqrt{3})/2+ma+na/2, \\ 
&&-(x\sqrt{3}+y)/2+na\sqrt{3}/2,z+(6k+4)b]\\
&&\rightarrow [(x+y\sqrt{3})/2+ma+na/2,\\
&&-(\sqrt{3}x-y)/2+na\sqrt{3}/2,z+(6k+5)b]
\end{eqnarray*}
According to the above the Casimir energy is given by 
\begin{eqnarray*}
&&\rho^c_{\mathbb{Z}_6}=-\frac{1}{\pi^2}\sum_{m,n,k=-\infty}^{+\infty}{}^\prime\left\{
\left[(ma+na/2)^2 +(na\sqrt{3}/2)^2\right.\right.\\ 
&&\left.\left.+(6kb)^2\right]^{-2}+\left[ (-(x/2+y\sqrt{3})/2+ma+na/2)^2 
\right. \right.\\
&&\left. \left.+( (\sqrt{3}x-y)/2+na\sqrt{3}/2 )^2 
+((6k+1)b)^2\right]^{-2}\right.\\
&&\left.  +\left [(-(3x+y\sqrt{3})/2+ma+na/2)^2 
\right. \right.\\
&&\left. \left.+( (\sqrt{3}x-3y)/2+na\sqrt{3}/2 )^2 
+((6k+2)b)^2\right]^{-2}\right.\\
&&\left.+\left[ (-2x +ma+na/2)^2 +(-2y+ na\sqrt{3}/2)^2\right.\right.\\
&& \left.\left. +((6k+3)b)^2\right]^{-2}
+\left[(-(3x-y\sqrt{3})/2+ma+na/2)^2 \right.\right. \\
&&\left.\left.+(-(\sqrt{3}x+3y)/2+na\sqrt{3}/2)^2 +((6k+4)b)^2\right]^{-2} 
\right.\\
&&\left. +\left[(-(x-y\sqrt{3})/2+ma+na/2)^2 \right.\right. \\
&&\left.\left.+(-(\sqrt{3}x+y)/2+na\sqrt{3}/2)^2 +((6k+5)b)^2\right]^{-2}
\right\}
\end{eqnarray*}
\subsection{$\Psi=\mathbb{Z}_2\otimes \mathbb{Z}_2$\label{ap4}}
\begin{eqnarray*}
&&\vec{r}=(x,y,z)\rightarrow (x+2ma,y+2nb,z+2kc) \\
&&\rightarrow (x+a+2ma,-y+2nb,-z+2kc)\\
&&\rightarrow (-x+2ma,y+b+2nb,-z+c+2kc)\\
&&\rightarrow (-x+a+2ma,-y+b+2nb,z+c+2kc)
\end{eqnarray*}
According to the above the Casimir energy is given by 
\begin{eqnarray*}
&&\rho^c_{\mathbb{Z}_2\otimes \mathbb{Z}_2}=-\frac{1}{\pi^2}
\sum_{m,n,k=-\infty}^{+\infty}{}^\prime\left\{
\left[(2ma)^2 +(2nb)^2+(2kc)^2\right]^{-2}\right.\\ 
&&\left.  +\left[(a+2ma)^2 +(-2y+2nb)^2 +(-2z-2kc)^2\right]^{-2}\right.\\
&&\left.+\left[(-2x+2ma)^2 +(b+2nb)^2 +(-2z+c+2kc)^2\right]^{-2} \right.\\
&&\left. +\left[(-2x+a+2ma)^2 +(-2y+b+2nb)^2 +(c+2kc)^2\right]^{-2} \right\}
\end{eqnarray*}

\end{document}